\documentclass[12pt]{article} 
\usepackage{bbold}
\usepackage{graphicx}
\def \b{{\cal B}} 
\def \bea{\begin{eqnarray}} 
\def \beq{\begin{equation}}

\def \eea{\end{eqnarray}} 
\def \eeq{\end{equation}}

\def \s{\sqrt{2}}

\def\lsim{\mathrel{\rlap{\lower3pt\hbox{$\sim$}}\raise2pt\hbox{$<$}}}
\def\gsim{\mathrel{\rlap{\lower3pt\hbox{$\sim$}}\raise2pt\hbox{$>$}}}

\begin{document} 
\begin{flushright}
EFI 11-16 \\
TECHNION-PH-2011-14 \\
July 2011 \\ 
\end{flushright} 
\centerline{\bf TRIPLE PRODUCT ASYMMETRIES IN $K, D_{(s)}$ AND $B_{(s)}$ DECAYS}
\medskip
\centerline{Michael Gronau} 
\centerline{\it Physics Department, Technion, Haifa 32000, Israel.}
\medskip 
\centerline{Jonathan L. Rosner} 
\centerline{\it Enrico Fermi Institute and Department of Physics,
  University of Chicago} 
\centerline{\it Chicago, IL 60637, U.S.A.} 
\bigskip 

\begin{quote}
One distinguishes between ``true" CP violating triple product (TP) 
asymmetries which require no strong phases and ``fake" asymmetries which are
due to strong phases but require no CP violation. So far a single true TP
asymmetry has been measured in $K_L\to \pi^+\pi^- e^+e^-$. A general discussion
is presented for T-odd TP asymmetries in four-body decays. It is shown that 
TP asymmetries vanish for two identical and kinematically indistinguishable 
particles in the final state.
Two examples are $D^0\to K^-\pi^+\pi^-\pi^+$ and $D^+\to K^-\pi^+\pi^+\pi^0$.
A non-zero TP asymmetry can be expected when non-trivial kinematic correlations
exist, as in the decay $K_L \to e^+ e^- e^+ e^-$.
Triple product asymmetries measured in charmed particle decays 
indicate an interesting pattern of final-state interactions.
We reiterate a discussion of TP asymmetries in 
$B$ meson decays to two vector mesons each decaying to a pseudoscalar 
pair, extending results to decays where one vector meson decays into 
a lepton pair. We derive expressions for time-dependent TP asymmetries
for neutral $B$ decays to flavorless states in terms of the neutral $B$ 
mass difference $\Delta m$ and the width-difference $\Delta\Gamma$.
Time-integrated true CP violating asymmetries, measurable 
for untagged $B_s$ decays, are shown to be suppressed by neither
$\Gamma_s/\Delta m_s$ nor $\Delta\Gamma_s/\Gamma_s$ if transversity amplitudes
for CP-even and CP-odd states involve different weak phases. 
In contrast, fake asymmetries require flavor tagging and are suppressed by 
the former ratio when time-integrated. 
We apply our results to $B\to K^*\phi$ and $B_s\to\phi\phi$ data and suggest 
an application for $B_s\to J/\psi\phi$.
\end{quote}

\leftline{\qquad PACS codes: 11.30.Er, 13.25.Hw, 14.40.Nd}

\section{Introduction}

A powerful tool for displaying CP violation in weak decays is the investigation
of triple product asymmetries \cite{Valencia:1988it,Kayser:1989vw,Datta:2003mj,%
Datta:2011qz}.  A four-body decay gives rise to three independent
final momenta in the rest frame of the decaying particle, and one can form
a T-odd expectation value out of (e.g.) $\vec{p}_1 \times \vec{p}_2
\cdot \vec{p}_3$.  Under certain circumstances a non-zero value of this
triple product can also signify CP violation.  A famous example is the
CP-odd asymmetry of $(13.6 \pm 1.4 \pm 1.5)\%$ reported by the KTeV
Collaboration \cite{AlaviHarati:1999ff}.  Here we present a general discussion
for T-odd triple product (TP) asymmetries in four-body decays of strange,
charmed, and beauty mesons, focusing on genuine CP violating asymmetries.
While these asymmetries are generally expected to be small in the Standard
Model, larger values can signify new physics, and their observation (in
contrast to direct CP asymmetries in decay rates) does not depend on the
presence of large (but generally incalculable) strong final-state phases.

Charmed meson decays are expected to exhibit very small CP violating
effects in the Standard Model~\cite{Bigi:2001sg}. Triple product asymmetries 
in four body $D$ and $D_s$ decays are expected to reflect final state
interactions.  Comparing triple product asymmetries in charmed meson decays and
in CP conjugate processes provides CP violating observables which could serve
as potential probes for new physics.

Focusing on $B$ meson decays, four-particle final states are obtained through 
two vector meson intermediate states.  Studying CP violating TP asymmetries is 
of particular interest in a class of decays which are induced by $b\to s$ 
transitions. These CKM (Cabibbo-Kobayashi-Maskawa) and loop suppressed 
processes are sensitive to new decay amplitudes~\cite{Grossman:1996ke}. 
$B_s$ decays to two vector mesons induced 
by $b\to c\bar cs$ involve in the Standard Model a very small weak phase 
occurring in the interference of $B_s$-$\bar B_s$ mixing and decay amplitudes.
This phase may be affected by new contributions to $B_s$-$\bar B_s$ mixing. 
The question is whether such new contributions could show up in TP asymmetries. 

In Section \ref{sec:TP} we lay the foundation for a discussion of triple
product asymmetries in four-body decays.  We specialize to an example of
neutral kaon decays in Sec.\ \ref{sec:K}.
Recently measured triple product asymmetries and CP violating asymmetries 
in charmed particle decays are discussed in Sec.\ \ref{sec:charm} drawing some 
conclusions about final state interactions.
A discussion of T-odd asymmetries is presented in Sec.\ \ref{sec:BVV}
for decays of a $B$ meson to a pair of vector mesons, which decay either to
two pseudoscalar pairs or to a pseudoscalar pair and a lepton pair.
The corresponding CP-violating TP asymmetries are then treated in Sec.\
\ref{sec:BVVCP}, studying time-dependence for asymmetries
in neutral $B$ decays in terms of a mass difference $\Delta m$ and a width
difference $\Delta\Gamma$. We discuss triple products for
specific $B$ decays to two vector mesons in Sec.~\ref{sec:TP2V} 
and present a short conclusion in Sec.\ \ref{sec:concl}.

\section{Triple products in four-body decays \label{sec:TP}} 

Scalar triple products (TP) of three-momentum or spin vectors occurring
in particle decays are interesting because they are odd under time-reversal T.
This may be due to a T-violating (and CP-violating) phase or caused by a
CP-conserving phase from final state interactions. A nontrivial triple product
requires at least four particles in the final state if only momenta are
measured.  Consider a four-body decay of a particle $P$, $P\to abcd$, in which
one measures the four
particles' momenta in the $P$ rest frame.  The momenta of the two pairs
of particles,  $ab$ and $cd$, form two decay planes intersecting at a straight
line given by the momentum vector $\vec p_a + \vec p_b = -\vec p_c - \vec p_d$.
We define $z$ to be the direction of $\vec p_a + \vec p_b$ and denote by $\hat
z$ a unit vector in this direction. Unit vectors normal to the two decay planes
and to their line of intersection $\hat z$ are denoted by $\hat n_{ab},
\hat n_{cd}$.  The angle $\phi$ between these two normal vectors is
conventionally defined to be the angle between the two decay planes. 

Thus we have
\beq
\hat n_{ab} \cdot \hat n_{cd} = \cos\phi~,~~~~
\hat n_{ab} \times \hat n_{cd} = \sin\phi \hat z~,
\eeq
implying a T-odd scalar triple product
\beq\label{sinphi}
(\hat n_{ab} \times \hat n_{cd})\cdot \hat z = \sin\phi~,
\eeq
and
\beq\label{sin2phi}
\sin 2\phi = 2(\hat n_{ab}\cdot \hat n_{cd})(\hat n_{ab}\times \hat n_{cd})
\cdot \hat z~,
\eeq
which is also odd under time-reversal because $\hat n_{ab}\cdot \hat n_{cd}$ is
even under this transformation.  A T-odd asymmetry in the decay can be defined
by an asymmetry between the number of events $N$ with positive and negative
values of $\sin\phi$ or $\sin 2\phi$: for example,
\beq\label{A_Ta}
A_T(\sin 2\phi) \equiv \frac{N(\sin 2\phi > 0) - N(\sin 2\phi < 0)}{N(\sin
2\phi > 0) + N(\sin 2\phi < 0)}~.
\eeq

A special example of this kind of asymmetry has been studied several years ago by the
KTeV and NA48 Collaborations in $K_L \to \pi^+\pi^-e^+e^-$, measuring values
$A_T(\sin 2\phi) = (13.6\pm 1.4\pm 1.5)\%$~\cite{AlaviHarati:1999ff} and
$A_T(\sin 2\phi) = (14.2 \pm 3.6)\%$~\cite{Lai:2003ad}, respectively. Here $\phi$
is the angle between vectors $\hat n_\pi$ and $\hat n_e$ which are normal to
the $\pi^+\pi^-$ and $e^+e^-$ planes, $\sin 2\phi = 2(\hat n_\pi \cdot \hat
n_e)(\hat n_\pi \times \hat n_e) \cdot \hat z$, $\hat z \equiv [\vec p(\pi^+) +
\vec p(\pi^-)]/|\vec p(\pi^+)+\vec p(\pi^-)|$. In this particular decay, which
involves two particle-antiparticle pairs, the quantity $\sin 2\phi$  changes
sign under both T and CP~\cite{Sehgal:1992wm}. The latter property can be
seen by noting that under C, $\vec p(\pi^\pm) \to \vec p(\pi^\mp),~ \vec
p(e^\pm) \to \vec p(e^\mp)$ while under P, $\vec p(\pi^\pm) \to - \vec
p(\pi^\pm), \vec p(e^\pm) \to -\vec p(e^\pm)$.
CP invariance would imply that the expectation value of this CP-odd observable 
vanishes for an initial CP-eigenstate.  Thus, {\it this measurement provides
the largest CP-nonconserving effect observed in kaon decays.} 

A particular case, in which the expectation value of a T-odd scalar triple
product of three momenta vanishes (irrespective of CP invariance) occurs when
two of the four final decay particles are identical, assuming that these
particles are kinematically indistinguishable. This happens when
one does not include a constraint on the final particle momenta. 
Two useful examples, which will
be discussed in Section \ref{sec:charm} with other charm decays, are 
$D^0\to K^-\pi^+\pi^-\pi^+$ and $D^+\to K^-\pi^+\pi^+\pi^0$ both of 
which involves two identical $\pi^+$ mesons in the final state.

A general proof of this property is based on the covariant form of a triple
product observable in $P\to abcd$ expressed as $\epsilon_{\mu\nu\rho\sigma}
p_a^\mu p_b^\nu p^\rho_c p^\sigma_d$ in terms of the four outgoing particle
four-momenta.  We are assuming that the final particles $a$ and $b$ are
identical and are kinematically indistinguishable.
Using energy-momentum conservation
($p_d = p_B -p_a -p_b - p_c$), the above expression becomes proportional to
$\epsilon_{ijk}p^i_a p^j_b p^k_c =(\vec p_a \times \vec p_b)\cdot \vec p_c=
-(\vec p_b \times \vec p_a) \cdot \vec p_c$ in the $B$ rest frame. Because of
its antisymmetry in $\vec p_a$ and $\vec p_b$, the expectation value of this
triple product vanishes, $\langle (\vec p_a\times \vec p_b)\cdot \vec p_c
\rangle=0$, when summing over the indistinguishable momenta of the two
identical particles. 

An alternative proof of this theorem for identical particles $a$ and $b$ may be
presented by showing that $A_T(\sin\phi)=0$ or $\langle \sin\phi\rangle = 0$,
where $\sin\phi$ is defined in Eq.~(\ref{sinphi}). Writing
\beq\label{sinphi2}
\sin\phi = \hat n_{ab} \cdot (\hat n_{cd} \times \hat z)~,
\eeq 
one has $\hat n_{ab} = (\vec p_a \times \vec p_b)/|\vec p_a \times \vec p_b|$
while $\hat n_{cd} \times \hat z$ is a vector in the plane of $\vec p_c$ and
$\vec p_d$ perpendicular to $\vec p_c + \vec p_d$. Using momentum conservation,
$\vec p_d = -\vec p_a -\vec p_b -\vec p_c$, the vector $\hat n_{cd} \times \hat
z$ may be replaced by $\vec p_c$ while $\vec p_a$ and $\vec p_b$ do not
contribute to (\ref{sinphi2}).  Thus 
\beq
\langle \sin\phi \rangle \propto \langle [(\vec p_a \times \vec p_b)\cdot \vec
p_c]/ |\vec p_a \times \vec p_b| \rangle~,
\eeq 
which vanishes when summing symmetrically over the momenta $\vec p_a$ and 
$\vec p_b$.

A nonzero triple product asymmetry may occur when at least one of the 
two identical particles forms a resonance, or favors a low invariant mass, 
with a third particle ($c$ or $d$), 
in which case one does not sum symmetrically over $\vec p_a$ and $\vec p_b$ in 
$\langle (\vec p_a \times \vec p_b) \cdot \vec p_c \rangle$.  In four-body
decays, where two pairs of final particles are associated with  two vector 
mesons in an intermediate state, the triple product asymmetry depends also on
the vector meson polarization and does not have to vanish for two identical 
particles.  This situation occurs in $B$ and $B_s$ decays to two vector mesons,
for instance in $B^0\to K^{*0}(\to K^+\pi^-)\phi(\to K^+K^-)$ 
and $B_s\to \phi(\to K^+K^-)\phi(\to K^+K^-)$.

\section{The decays $K_L \to e^+e^-e^+e^-$ and $K_L \to e^+e^-\mu^+\mu^-$
\label{sec:K}}

A simple example demonstrates the above circumstances permitting a CP- or
T-violating expectation value in a four-body decay even when two pairs of
final-state particles are equal.  This is in the decay $K_L \to e^+e^-e^+e^-$
for which 441 and 200 events were observed by the KTeV
\cite{AlaviHarati:2001ab} and NA48~\cite{Lai:2005kw} collaborations. (The decay
$K_L \to e^+e^-\mu^+\mu^-$ also has been observed 
by KTeV~\cite{AlaviHarati:2001tk}.)
Consider first of all only very low-mass $e^+ e^-$ pairs produced by photons
very near their mass shell.  

Define the CP-even and CP-odd combinations of
$K^0$ and $\bar K^0$ to be $K_1$ and $K_2$, respectively.  We have $K_L \simeq
K_2 + \epsilon K_1$, where $|\epsilon| = (2.228 \pm 0.011) \times 10^{-3}$,
Arg$(\epsilon) = (43.51 \pm 0.05)^\circ$~\cite{Nakamura:2010zzi}.  
Since the $K_L$ is mainly CP-odd, its decay to
two photons is dominated by the effective Lagrangian ${\cal L}_- \propto
K_2 F_{\mu \nu} \tilde F^{\mu \nu}$, but the small CP-even admixture decays
via an effective Lagrangian ${\cal L}_+ \propto K_1 F_{\mu \nu} F^{\mu \nu}$.
Here
\beq
F_{\mu \nu} = \left[ \begin{array}{c c c c} 0 & E_1 & E_2 & E_3 \cr
-E_1 & 0 & - B_3 & B_2 \cr -E_2 & B_3 & 0 & -B_1 \cr -E_3 & -B_2 & B_1 & 0
\end{array} \right]~,~~
\tilde F_{\mu \nu} = \left[ \begin{array}{c c c c} 0 & -B_1 & -B_2 & -B_3 \cr
B_1 & 0 & -E_3 & E_2 \cr B_2 & E_3 & 0 & -E_1 \cr B_3 & -E_2 & E_1 & 0
\end{array} \right]~,
\eeq
so that ${\cal L}_+ \propto K_1(\vec{B}^2-\vec{E}^2)$, ${\cal L}_- \propto 2K_2
\vec{E} \cdot \vec{B}$.  Let one photon be emitted along the $+\hat z$ axis
with polarization $\epsilon_1 = \hat x$, and measure the polarization of a
second photon along
the $-\hat z$ axis with a polarizer oriented in the direction $\epsilon_2 =
\hat x \cos \phi + \hat y \sin \phi$.  For the decay of a CP-(even,odd)
state, the amplitudes for observing this photon are then proportional
to $\cos \phi,~\sin \phi$, respectively \cite{Yang:1950rg}.  The decay of a
CP admixture such as $K_L$ then will give rise to interference between these
two amplitudes and hence an amplitude proportional to $\sin(\phi - \delta)$,
where $\delta \ne (0,\pi/2)$.

In the case of $K_L \to e^+ e^- e^+ e^-$, the virtual photons giving rise
to $e^+ e^-$ pairs are not exclusively transversely polarized, and the
$e^+ e^-$ planes do not analyze photon polarizations perfectly, so that the
signal for even or odd CP will be diluted.  For example, in the case of
$\pi^0 \to e^+ e^- e^+ e^-$ \cite{Kroll:1955zu}, the angular distribution
of the decay rate is
\beq
\pi \frac{1}{\Gamma}\frac{d \Gamma}{d \phi} = (0.59 \sin^2 \phi
+ 0.41 \cos^2 \phi)~.
\eeq
whereas an argument based on transversely polarized photons would have given
$\sin^2 \phi$ for the right-hand side.  For $K_L \to e^+ e^- e^+ e^-$ one finds 
assuming no direct CP violation~\cite{Kroll:1955zu,Gu:1994nb}
\beq\label{eq:dG/dP}
2 \pi \frac{1}{\Gamma}\frac{d \Gamma}{d \phi} = 1 + \beta_{CP} \cos(2\phi) +
\gamma_{CP} \sin(2 \phi)~,
\eeq
\beq \label{eq:bg}
\beta_{CP} \equiv \frac{1 - |\epsilon r|^2}{1 + |\epsilon r|^2} B~,~~
\gamma_{CP} \equiv \frac{2 {\rm Re}(\epsilon r)}{1 + |\epsilon r|^2} C~,
\eeq
where $r \equiv |A(K_1 \to e^+ e^- e^+ e^-)/A(K_2 \to e^+ e^- e^+ e^-)|$ is
of order unity, $B \simeq -0.2$ (it would be +0.2 for $K_S \to e^+ e^- e^+
e^-$), and $C$ has not yet been calculated.
One would expect $C$ to be of the same order as $B$ as it represents a
``dilution'' of the interference between CP-even and CP-odd decays as analyzed
by the electron-positron pairs.  

The term $\gamma_{CP}$ is directly related to the
T-odd observable in Eq.\ (\ref{A_Ta}),
\beq
A_T(\sin 2 \phi) = (2/\pi) \gamma_{CP}~,
\eeq
which in this case of two particle-antiparticle pairs in the final state is
also CP-odd.  Measured values of $\beta_{CP}$ and $\gamma_{CP}$ are shown in
Table \ref{tab:bg}.  They are consistent with theoretical predictions,
although improvement of accuracy by at least a factor of 100 will be needed
to see nonzero $\gamma_{CP}$ at the predicted level.  We thus show that in
order to form a T and CP-odd observable it is not necessary to have four
distinct particles as long as they exhibit non-trivial kinematic correlations.

\begin{table}
\caption{Measured values of $\beta_{CP}$ and $\gamma_{CP}$ [Eqs.\ 
(\ref{eq:dG/dP})(\ref{eq:bg})] in $K_L \to e^+ e^- e^+ e^-$.
\label{tab:bg}}
\begin{center}
\begin{tabular}{c c c} \hline \hline
Collaboration & KTeV \cite{AlaviHarati:2001ab} & NA48 \cite{Lai:2005kw} \\
\hline
Events & 441 & 200 \\
$\beta_{CP}$ & $-0.23 \pm 0.09 \pm 0.02$ & $-0.13 \pm 0.10 \pm 0.03$ \\
$\gamma_{CP}$ & $-0.09 \pm 0.09 \pm 0.02$ & $0.13 \pm 0.10 \pm 0.03$ \\
\hline \hline
\end{tabular}
\end{center}
\end{table}

\section{TP and CP violating asymmetries
in $D_{(s)}$ decays \label{sec:charm}}

Four-body Cabibbo-favored $D$ and $D_s$ decays involve sizable 
branching ratios. For instance, a few years ago the CLEO collaboration reported 
measurements~\cite{:2007zt} ${\cal B}(D^0\to K^-\pi^+\pi^-\pi^+) 
= (8.30\pm 0.07\pm 0.20)\%$, ${\cal B}(D^+\to K^-\pi^+\pi^+\pi^0)=
(5.98\pm 0.08\pm 0.18)\%$ and~\cite{:2008cqa} 
${\cal B}(D_s \to K^+K^-\pi^+\pi^0)=(5.65\pm 0.29\pm 0.40)\%$.
As we have shown in Section \ref{sec:TP}, triple product asymmetries 
are expected to vanish in the first two processes both of which involve 
two identical $\pi^+$ mesons which are kinematically indistinguishable. 

Triple product correlations have been studied by the FOCUS and BaBar
collaborations in Cabibbo-suppressed decays $D^0\to K^+K^-\pi^+\pi^-$
\cite{Link:2005th,delAmoSanchez:2010xj} and very recently by the BaBar
collaboration in both Cabibbo-favored and Cabibbo-suppressed decays, $D^+_s\to
K^+K_S\pi^+\pi^-$  and $D^+\to K^+K_S\pi^+\pi^-$, respectively~\cite{:2011dx}. 
Denoting a scalar triple product for momenta of three final particles
in the charmed meson rest frame, 
$C_T\equiv  \vec p_1\cdot(\vec p_2\times\vec p_3)$, one defines a 
triple product asymmetry for $D$ or $D_s$ decay~\cite{Bigi:2001sg}
\beq  
A_T \equiv \frac{\Gamma(C_T>0) - \Gamma(C_T<0)}
{\Gamma(C_T>0) + \Gamma(C_T<0)}~.
\eeq
This T-odd asymmetry is expected to be nonzero as a result of final state 
interactions. In order to test for CP violation one compares this asymmetry with 
a corresponding asymmetry in the CP conjugate process involving $\bar D$ 
or $\bar D_s$,
\beq
\bar A_T \equiv \frac{\Gamma(-\bar C_T>0) - \Gamma(-\bar C_T<0)}
{\Gamma(-\bar C_T>0) + \Gamma(-\bar C_T<0)}~.
 \eeq
 Here $\bar C_T$ denotes a triple product of momenta for charge-conjugate 
 particles while the minus sign in front of $\bar C_T$ follows by applying parity.  
 
 The difference
 \beq
 {\cal A}_T \equiv \frac{1}{2}(A_T - \bar A_T)
 \eeq
 provides a measure for CP violation. A nonzero asymmetry ${\cal A}_T$ may 
 follow from a CP asymmetry in partial rates. In the absence of
 such asymmetry [assuming $\Gamma(-\bar C_T>0) + \Gamma(-\bar C_T<0)=
 \Gamma(C_T>0) + \Gamma(C_T<0)$]
  ${\cal A}_T\ne 0$ may be the result of a CP asymmetry
 in triple product correlations, $\Gamma(-\bar C_T>0) - \Gamma(-\bar C_T<0)
 \ne \Gamma(C_T>0) - \Gamma(C_T<0)$.   
 
\begin{table}
\caption{Triple-product asymmetries $A_T, \bar A_T, {\cal A}_T$ and $\Sigma_T$
(defined in the text) for Cabibbo-suppressed decays $D^0 \to K^+K^- \pi^+
\pi^-$~\cite{delAmoSanchez:2010xj}, $D^+ \to K^+ K_S \pi^+ \pi^-$
\cite{:2011dx} and Cabibbo-favored decays $D_s^+ \to K^+ K_S \pi^+\pi^-$
\cite{:2011dx}.  Values are quoted in units of $10^{-3}$.
\label{tab:A2}}
\begin{center}
\begin{tabular}{c c c c} \hline \hline
Asymmetry & $D^0/\bar D^0$ & $D^+/D^-$ & $D_s^+/D_s^-$ \\ \hline
  $A_T$    & $-68.5\pm7.3\pm5.8$ & $11.2\pm14.1\pm5.7$ &
 $-99.2\pm10.7\pm8.3$ \\
$\bar A_T$ & $-70.5\pm7.3\pm3.9$ & $35.1\pm14.3\pm7.2$ &
 $-72.1\pm10.9\pm10.7$ \\
 ${\cal A_T}$ & $1.0\pm 5.1\pm 4.4$ & $-12.0\pm 10.0\pm 4.6$ & 
 $-13.6 \pm 7.7 \pm 3.4$ \\
 $\Sigma_T$ & $-69.5\pm 6.2$ & $23.1\pm 11.0$ & $85.6 \pm 10.2$
  \\ \hline \hline
\end{tabular}
\end{center}
\end{table}
 
Table \ref{tab:A2} quotes values of $A_T, \bar A_T$ and ${\cal A}_T$ 
 from Refs.~\cite{delAmoSanchez:2010xj,:2011dx} for Cabibbo-suppressed
$D^0\to K^+K^-\pi^+\pi^-$, $D^+\to K^+ K_S\pi^+\pi^-$ and Cabibbo-favored
$D^+_s\to K^+K_S\pi^+\pi^-$. For completeness we also
include in the table values calculated for a quantity
\beq
\Sigma_T \equiv \frac{1}{2}(A_T + \bar A_T)~.
\eeq 
This average of triple product asymmetries in a charmed meson decay and its
CP conjugate is not CP violating. Rather, being T-odd, it may provide
information on final state interaction.

While all three values of ${\cal A}_T$ in Table \ref{tab:A2} are consistent
with zero, the values of $\Sigma_T$ are considerably more significant for
$D^0$ and $D_s^+$ decays than for $D^+$ decays.  This pattern seems to indicate
a difference among final-state interactions in the three decays.  Final-state
interactions
in Cabibbo-favored $D$ decays could in part be responsible for the hierarchy of
lifetimes $\tau(D^+) > \tau(D_s^+) \gsim \tau(D^0)$.  The final states in
Cabibbo-favored $D^+$ decays are ``exotic'' involving $I=3/2$ with quantum
numbers of $s u \bar d \bar d$, and do not correspond to any known resonances,
whereas Cabibbo-favored $D^0$ and $D_s^+$ decays populate $I=1/2$ and $I=1$
states with quantum numbers of $s \bar d$ and $u \bar d$, respectively.
The measured longer $D^+$ lifetime could thus be associated with the lack of
resonances contributing to its decays \cite{Gaillard:1974mw,Gronau:1999zt}.  

One may perhaps expect an enhancement pattern similar to the one observed 
in the total hadronic decay rate of $D^0$ relative to that of $D^+$ also in 
Cabibbo-suppressed decays.
The total hadronic enhancement is given by~\cite{Nakamura:2010zzi}
\beq\label{had-enhance}
\frac{\Gamma_h(D^0)}{\Gamma_h(D^+)}
= \frac{\tau(D^+)}{\tau(D^0)}\left(\frac{1-\b_{\rm sl}(D^0)}{1-\b_{\rm sl}(D^+)}\right)= 
\frac{1040\pm 7}{410.1 \pm 1.5}\left(\frac{0.868\pm 0.006}{0.66\pm 0.03}\right)
=3.34 \pm 0.15~.
\eeq
Here $\b_{\rm sl}\equiv \b_{{\rm sl},e} + \b_{{\rm sl},\mu}$ are
semileptonic branching ratios, 
$\b_{{\rm sl},e}(D^0) = (6.49\pm 0.11)\%$, $\b_{{\rm sl},\mu}(D^0) 
=(6.7\pm 0.6)\%$, $\b_{{\rm sl},e}(D^+) = (16.07\pm 0.30)\%$, $\b_{{\rm sl},\mu}(D^+) 
=(17.6\pm 3.2)\%$.
Using~\cite{Nakamura:2010zzi}
$\b(D^0\to K^+K^-\pi^+\pi^-)=(2.42\pm 0.12)\times 10^{-3},
\b(D^+\to K^+K_S\pi^+\pi^-) =(1.75\pm 0.18)\times 10^{-3}$, one calculates the 
ratio of Cabibbo-suppressed decay rates,
\beq\label{CSratio}
\frac{\Gamma(D^0\to K^+K^-\pi^+\pi^-)}{\Gamma(D^+\to K^+\bar K^0\pi^+\pi^-)}
=\frac{\tau(D^+)}{\tau(D^0)}
\frac{\b(D^0\to K^+K^-\pi^+\pi^-)}{2\b(D^+\to K^+ K_S\pi^+\pi^-)}
= 1.75 \pm 0.20~.
\eeq
Thus we conclude that some enhancement of Cabibbo-suppressed 
$D^0\to K^+K^-\pi^+\pi^-$ relative to $D^+\to K^+\bar K^0\pi^+\pi^-$ occurs, 
although it is less than in Cabibbo-favored decays. 

This partial enhancement may account for the pattern of measured values of 
$\Sigma_T$ quoted for these two Cabibbo-suppressed processes in Table 
\ref{tab:A2}. The large value of $\Sigma_T$ measured for 
$D^+_s\to K^+K_S\pi^+\pi^-$ reflects an enhancement in $D^+_s$ 
Cabibbo-favored decay rates. A total hadronic enhancement factor for $D^+_s$ 
similar to (\ref{had-enhance}), $\Gamma_h(D^+_s)/\Gamma_h(D^+) \simeq 2.6$, 
is calculated including in the numerator a subtraction of
$\b(D^+_s \to \tau^+\nu_\tau)=(5.43\pm 0.31)\%$~\cite{Nakamura:2010zzi}.

\section{T-odd asymmetries in $B_{(s)} \to V_1V_2$ \label{sec:BVV}} 

Consider $B_{(s)}$ decays into two vector mesons $V_1$ and $V_2$, each decaying
to a pair of pseudoscalars, $P_1P'_1$ and $P_2P'_2$. The decay amplitude for
$B_{(s)}(p) \to V_1(k_1,\epsilon_1) + V_2(k_2,\epsilon_2)$ may be written in
terms of angular momentum amplitudes~\cite{Valencia:1988it} (we use 
normalization as in~\cite{Datta:2003mj}),
\beq
M = {\bf a}\epsilon^*_1 \cdot \epsilon^*_2 + 
\frac{{\bf b}}{m^2_B}(p \cdot \epsilon^*_1)(p \cdot \epsilon^*_2)
+i\frac{{\bf c}}{m^2_B}\epsilon_{\mu\nu\rho\sigma} p^\mu q^\nu
\epsilon^{*\rho}_1 \epsilon^{*\sigma}_2~,
\eeq 
where $q\equiv k_1-k_2$. The amplitudes ${\bf a}$ and ${\bf b}$ are linear 
combinations of $S$ and $D$ wave amplitudes while ${\bf c}$ corresponds to $P$
wave.  It is customary to use transversity amplitudes~\cite{Dighe:1995pd},
which are related to the angular momentum amplitudes through the following 
relations~\cite{Datta:2003mj} (see also~\cite{Kramer:1991xw} for relations
involving helicity amplitudes):
\beq
A_\parallel = \s {\bf a}~,~~~~A_0 = -{\bf a}x - \frac{m_1m_2}{m^2_B} {\bf b}
(x^2 - 1)~,~~~ A_\perp = 2\s\frac{m_1m_2}{m^2_B}{\bf c}\sqrt{x^2-1}~.
\eeq
Here $x \equiv (k_1\cdot k_2)/(m_1m_2); m_1$ and $m_2$ are the masses 
of $V_1$ and $V_2$.

\subsection{$V_1 \to P_1P'_1, V_2 \to P_2P'_2$} \label{subsec:4P}

Let us consider decays in which each of the two vector mesons in $B_{(s)}\to
V_1V_2$ decays into two pseudoscalar mesons. This class of decays consists of
charmless decays of $B$ and $B_s$ mesons including $B \to \phi(\to K^+K^-)
K^*(\to K\pi)$ and $B_s \to \phi(\to K^+ K^-)\phi(\to K^+K^-)$.  We denote by 
$\theta_1$ ($\theta_2$) the angle between the directions of motion of $P_1$
($P_2$) in the $V_1$ ($V_2$) rest frame and $V_1$($V_2$) in the $B$ rest frame.
The angle between the planes defined by $P_1P'_1$ and $P_2P'_2$ in the $B_{(s)}$ rest
frame will be denoted by $\phi$ as in Section \ref{sec:TP}.
The decay angular distribution in these three angles is given in terms of the
three transversity amplitudes $A_0, A_\parallel, A_\perp$~\cite{Sinha:1997zu}
(see also \cite{Kramer:1991xw}):
\bea\label{angular}
\frac{d\Gamma}{d\cos\theta_1d\cos\theta_2 d\phi} &=& 
N\left(|A_0|^2\cos^2\theta_1\cos^2\theta_2 + 
\frac{|A_\parallel|^2}{2}\sin^2\theta_1\sin^2\theta_2\cos^2\phi \right. 
 \\ 
&+& \frac{|A_\perp|^2}{2}\sin^2\theta_1\sin^2\theta_2\sin^2\phi 
+\frac{{\rm Re}(A_0A^*_\parallel)}{2\s}\sin 2\theta_1\sin 2\theta_2\cos\phi 
\nonumber 
\eea
\vskip -6mm
$$
~~~~~~~~~~~~~~~~~~~
\left. -~\frac{{\rm Im}(A_\perp A^*_0)}{2\s}\sin 2\theta_1\sin 2\theta_2
\sin\phi -  \frac{{\rm Im}(A_\perp A^*_\parallel)}{2}\sin^2\theta_1
\sin^2\theta_2\sin 2\phi \right)~.
\nonumber
$$
Integrating over $\theta_1$ and $\theta_2$ and using
\beq
\int^1_{-1}\cos^2\theta~d\cos\theta = \frac{2}{3}~,~~
\int^1_{-1}\sin^2\theta~d\cos\theta = \frac{4}{3}~,
\int^1_{-1}\sin 2\theta~d\cos\theta = 0~,
\eeq
one obtains the following distribution in $\phi$:
\beq\label{dG/dp}
\frac{d\Gamma}{d\phi} = \frac{4}{9}N\left(|A_0|^2 + 2|A_\perp|^2\sin^2\phi + 
2|A_\parallel|^2\cos^2\phi-2{\rm Im}(A_\perp A^*_\parallel)\sin 2\phi\right)~.
\eeq
The last term in this angular distribution provides a potential T-odd
asymmetry. Note that the term involving ${\rm Im}(A_\perp A^*_0)$ does not
contribute to a T-odd asymmetry when integrating over the angle $\theta_1$ or
$\theta_2$. 

One has now in analogy with Eqs.~(\ref{sinphi}) and (\ref{sin2phi}),
\beq\label{TP}
\sin\phi = (\hat n_{V_1}\times \hat n_{V_2})\cdot \hat p_{V_1}~,~~~~
\sin 2\phi = 2(\hat n_{V_1}\cdot\hat n_{V_2}) 
(\hat n_{V_1}\times \hat n_{V_2})\cdot \hat p_{V_1}~,
\eeq
where $\hat n_{V_i} (i=1,2)$ is a unit vector perpendicular to the $V_i$ decay
plane and $\hat p_{V_1}$ is a unit vector in the direction of $V_1$ in the $B_{(s)}$
rest frame.  A triple product (or more precisely a T-odd) asymmetry is now
defined similarly to Eq.~(\ref{A_Ta}) as an asymmetry between the number of
decays involving positive and negative values of $\sin
2\phi$~\cite{Datta:2003mj}:
\bea\label{A2_T}
A^{(2)}_T &  \equiv & \frac{\Gamma(\sin 2\phi >0) - \Gamma(\sin 2\phi<0)} 
{\Gamma(\sin 2\phi >0) + \Gamma(\sin 2\phi<0)} 
\nonumber\\
& = & \frac{[\int_0^{\pi/2}+ \int_\pi^{3\pi/2}](d\Gamma/d\phi)d\phi - 
[\int_{\pi/2}^\pi+\int_{3\pi/2}^{2\pi}] (d\Gamma/d\phi)d\phi}
{\int_0^{2\pi} (d\Gamma/d\phi)d\phi}~.
\eea
Using (\ref{dG/dp}) one obtains
\beq\label{expA2_T}
A^{(2)}_T = -\frac{4}{\pi}\frac{{\rm Im}(A_\perp A^*_\parallel)}{|A_0|^2 
+ |A_\perp|^2 + |A_\parallel|^2}~.
\eeq

The dependence of the angular distribution (\ref{angular}) on $\theta_1$ and
$\theta_2$ permits considering a second triple product 
asymmetry~\cite{Datta:2003mj} (or, more
precisely, a $T$-odd asymmetry) $A^{(1)}_T$ involving the ratio ${\rm Im}
(A_\perp A^*_0)/(|A_0|^2 + |A_\perp|^2 + |A_\parallel|^2)$.  One defines an
asymmetry with respect to values of $\sin\phi$ (a triple product), assigning it
the sign of $\cos\theta_1\cos\theta_2$ (a T-even quantity) and integrating over
all angles, 
\beq\label{A1_T}
A^{(1)}_T \equiv \frac{\Gamma[{\rm sign}(\cos\theta_1\cos\theta_2)\sin\phi > 0]
- \Gamma[{\rm sign}(\cos\theta_1\cos\theta_2)\sin\phi < 0]}
{\Gamma[{\rm sign}(\cos\theta_1\cos\theta_2)\sin\phi > 0]
+ \Gamma[{\rm sign}(\cos\theta_1\cos\theta_2)\sin\phi < 0]}~.
\eeq
A straightforward calculation using Eq.~(\ref{angular}) gives
 \beq\label{expA1_T}
 A^{(1)}_T = -\frac{2\s}{\pi}\frac{{\rm Im}(A_\perp A^*_0)}{|A_0|^2 +
 |A_\perp|^2 + |A_\parallel|^2}~.
 \eeq
 
The two triple product asymmetries, defined in Eqs.~(\ref{A2_T}) and
(\ref{A1_T}) and given in (\ref{expA2_T}) and (\ref{expA1_T}) in terms of
transversity amplitudes, are odd under time-reversal; however, they are not
genuine CP-violating or T-violating observables. Rather, they may be nonzero
due to a CP-conserving phase difference between two corresponding transversity
amplitudes while the weak phase difference of these amplitudes vanishes.
 
\subsection{$V_1 \to P_1P'_1, V_2 \to \ell^+\ell^-$} \label{subsec:2P2l}
 
We now consider a second class of decays into two vector mesons of which one 
meson decays into a pair of pseudoscalars while the other decays into a 
lepton pair $\ell^+\ell^-$, $\ell = e, \mu$. This class of processes involving
charmonium in the final state includes the decays $B  \to K^*(\to K\pi)J/\psi
(\to \mu^+\mu^-)$ and $B_s \to \phi(\to K^+K^-)J/\psi(\to \mu^+\mu^-)$.
As in decays into four pseudoscalars, we denote by 
$\theta_1$ the angle between the directions of motion of $P_1$
in the $V_1$ rest frame and $V_1$ in the $B_{(s)}$ rest frame, while $\theta_\ell$
is the corresponding angle of $\ell^+$ in the $V_2$ rest frame. The angle 
between the planes defined by $P_1P'_1$ and $\ell^+\ell^-$ in the $B_{(s)}$ rest 
frame will be denoted here by $\phi$.  One is interested in triple products
which are functions of this angle.

The complete decay angular distribution for this class of 
decays is given by~\cite{Dighe:1995pd} (see also~\cite{Kramer:1991xw}):
\bea \label{angular'}
\frac{d\Gamma}{d\cos\theta_1d\cos\theta_\ell d\phi} &=& 
N\left(|A^\ell_0|^2\cos^2\theta_1\sin^2\theta_\ell + 
\frac{|A^\ell_\parallel|^2}{2}\sin^2\theta_1 (\sin^2\phi +
\cos^2\theta_\ell\cos^2\phi) \right. 
\nonumber \\ 
& + & \frac{|A^\ell_\perp|^2}{2} \sin^2\theta_1 (\cos^2\phi+
\cos^2\theta_\ell\sin^2\phi) \\
& + & \frac{1}{2\s} {\rm Im}(A^\ell_\perp A^{\ell*}_0) \sin 2 \theta_1 \sin
2 \theta_\ell \sin\phi
 \nonumber \\
& - & \frac{{\rm Re}(A^\ell_0A^{\ell*}_\parallel)}{2\s}\sin 2\theta_1
\sin 2\theta_\ell\cos\phi 
\nonumber \\
& + & \left. \frac{1}{2} {\rm Im}(A^\ell_\perp A^{\ell*}_\parallel)
\sin^2 \theta_1 \sin^2 \theta_2 \sin 2\phi \right)~.
\eea
Integrating over the angles $\theta_1$ and $\theta_\ell$ one obtains:
\bea
\frac{d\Gamma}{d\phi} & = & \frac{4}{9}N\left(2|A^\ell_0|^2 + |A^\ell_\parallel|^2(1 + 2\sin^2\phi)
+ |A^\ell_\perp|^2(1+2\cos^2\phi) \right.
\nonumber\\
& + & \left. 2\,{\rm Im}(A^\ell_\perp A^{\ell*}_\parallel)\sin 2\phi\right)~.
\eea
The last term is a source of one of two triple product asymmetries.  A T-odd
asymmetry defined for $\sin 2\phi$ in analogy with (\ref{A2_T}) obtains a
similar expression (but different 
sign and normalization) in terms of transversity
amplitudes,
\beq\label{A2_Tl}
A_T^{(2)\ell}   \equiv  \frac{\Gamma(\sin 2\phi >0) - \Gamma(\sin 2\phi<0)} 
{\Gamma(\sin 2\phi >0) + \Gamma(\sin 2\phi<0)} 
=  \frac{2}{\pi}\frac{{\rm Im}(A^\ell_\perp A^{\ell*}_\parallel)}{|A^\ell_0|^2
+ |A^\ell_\perp|^2 + |A^\ell_\parallel|^2}~.
\eeq
A second asymmetry can be defined for values of the triple product $\sin\phi$,
in the same manner as Eq.\ (\ref{A1_T}).  One obtains:
\beq\label{A1_Tl}
A^{(1)\ell}_T 
=  \frac{\s}{\pi}\frac{{\rm Im}(A^\ell_\perp A^{\ell*}_0)}{|A_0|^2 + 
|A^\ell_\perp|^2 + |A^\ell_\parallel|^2}~.
\eeq

\section{CP-violating TP asymmetries in $B_{(s)}\to V_1V_2$ \label{sec:BVVCP}}

\subsection{Self-tagged decays of charged and neutral $B$ mesons}
\label{subsec:self-tag}

In this subsection we consider $B$ and $B_s$ decays to states with specific
flavor, e.g. $B^{(+,0)}\to K^{*(+,0)}\phi$ and $B^{(+,0)} \to K^{*(+,0)}J/\psi$
belonging to the two classes considered in subsections \ref{subsec:4P} and
\ref{subsec:2P2l}, respectively.  We denote by $\bar A_0, \bar A_\parallel$ and
$\bar A_\perp$ transversity amplitudes for the CP-conjugate decay $\bar B_{(s)} \to
\bar V_1 \bar V_2$. The corresponding three angles describing the two vector
meson decays into pairs of pseudoscalar mesons will be denoted by
$\bar\theta_1, \bar\theta_2$ and $\bar\phi$. The decay angular distribution for
$\bar B_{(s)}$ decays has an expression similar to $B_{(s)}$ decays. The two 
terms linear in the parity-odd amplitude $\bar A_\perp$ change sign relative to the
corresponding two terms in Eq.~(\ref{angular}). Thus, for decays in which both
vector mesons $\bar V_1$ and $\bar V_2$ decay to a pseudoscalar pair one has:
\bea\label{angularBbar}
\frac{d\bar \Gamma}{d\cos\bar\theta_1d\cos\bar\theta_2 d\bar\phi} &=& 
N\left(|\bar A_0|^2\cos^2\bar\theta_1\cos^2\bar\theta_2 + 
\frac{|\bar A_\perp|^2}{2}\sin^2\bar\theta_1\sin^2\bar\theta_2\sin^2\bar\phi
 \right.  \\ 
&+& \frac{|\bar A_\parallel|^2}{2}\sin^2\bar\theta_1\sin^2\bar\theta_2
\cos^2\bar\phi +\frac{{\rm Re}(\bar A_0\bar A^*_\parallel)}{2\s}
\sin 2\bar\theta_1\sin 2\bar\theta_2\cos\bar\phi 
\nonumber 
\eea
\vskip -6mm
$$
~~~~~~~~~~~~~~~~~~~
\left. +~\frac{{\rm Im}(\bar A_\perp \bar A^*_0)}{2\s}\sin 2\bar \theta_1
\sin 2\bar \theta_2\sin\bar\phi +  \frac{{\rm Im}(\bar A_\perp \bar
A^*_\parallel)}{2}\sin^2\bar \theta_1\sin^2\bar\theta_2\sin 2\bar\phi \right)~.
\nonumber
$$

It has been pointed out~\cite{Valencia:1988it,Datta:2003mj} that the two
quantities ${\rm Im}(A_\perp A^*_0 - \bar A_\perp\bar A^*_0)$ and ${\rm Im}
(A_\perp A^*_\parallel - \bar A_\perp\bar A^*_\parallel)$, {\em occurring in
the sum (rather than the difference) of decay distributions (\ref{angular}) and
(\ref{angularBbar}) for $B_{(s)}$ and $\bar B_{(s)}$} for $\bar \theta_1=
\theta_1,\bar\theta_2=\theta_2,\bar\phi=\phi$, {\em are genuinely CP-violating}
and do not require nonzero CP conserving phases.  For instance, assuming that
each of the transversity amplitudes is dominated by a magnitude, $|A_\lambda|$,
a single CP-conserving phase, $\delta_\lambda$, and a single CP-violating phase,
$\phi_\lambda$ (which amounts to assuming no direct CP violation),
\beq
A_\lambda = |A_\lambda| e^{i\delta_\lambda} e^{i\phi_\lambda}~,~~~
\bar A_\lambda = |A_\lambda| e^{i\delta_\lambda} e^{-i\phi_\lambda}~~~
(\lambda=0,\parallel,\perp)~,
\eeq
implies
\beq\label{true}
{\rm Im}(A_\perp A^*_0 - \bar A_\perp\bar A^*_0) = 
2|A_\perp||A_0|\cos(\delta_\perp - \delta_0)\sin(\phi_\perp - \phi_0)~.
\eeq
This ``true" CP violating quantity is nonzero also when the CP-conserving phase
difference $\delta_\perp - \delta_0$ vanishes, provided that the CP-violating
phase difference $\phi_\perp -\phi_0$ between the two transversity amplitudes
$A_\perp$ and $A_0$ is nonzero.  In contrast, a quantity {\em occurring in the
difference of rates for $B_{(s)}$ and $\bar B_{(s)}$}, 
\beq\label{fake}
{\rm Im}(A_\perp A^*_0 + \bar A_\perp\bar A^*_0) = 
2|A_\perp||A_0|\sin(\delta_\perp - \delta_0)\cos(\phi_\perp - \phi_0)~,
\eeq
{\em is not CP-violating} as it is nonzero also when CP-violating phases
vanish.  Such a quantity will sometimes be referred to as a ``fake" asymmetry.

It is interesting to note that the CP-violating quantities 
${\rm Im}(A_\perp A^*_0 - \bar A_\perp\bar A^*_0)$ and 
${\rm Im}(A_\perp A^*_\parallel - \bar A_\perp\bar A^*_\parallel)$ occur in  
{\it triple product asymmetries for CP-averaged decay rates}. 
We denote partial decay rates for $B_{(s)}\to f$ and $\bar B_{(s)}\to \bar f$ 
by $\Gamma(B_{(s)}\to f)$ and  $\bar\Gamma(\bar B_{(s)}\to\bar f)$, respectively.  
The charge-averaged decay rate is  $[\Gamma(B_{(s)}\to f) + \bar\Gamma(\bar 
B_{(s)}\to\bar f)]/2$, and a triple
product asymmetry defined for this rate is given by:
\bea\label{A2cave}
{\cal A}_T^{(2) \rm chg-avg} & \equiv & \frac{[\Gamma(\sin 2\phi >0) +
\bar\Gamma(\sin 2\bar\phi>0)] - [ \Gamma(\sin 2\phi<0) + \bar\Gamma(\sin 2\bar
\phi<0)]} {[\Gamma(\sin 2\phi >0) + \bar\Gamma(\sin 2\bar\phi>0)]  +
[ \Gamma(\sin 2\phi<0) + \bar\Gamma(\sin 2\bar \phi<0)]} \nonumber\\
& = & -\frac{4}{\pi}\frac{{\rm Im}(A_\perp A^*_\parallel - \bar A_\perp\bar
A^*_\parallel)} {(|A_0|^2 + |A_\perp|^2 + |A_\parallel|^2) + (|\bar A_0|^2 +
|\bar A_\perp|^2 + |\bar A_\parallel|^2)}~.
\eea 
As noted above the numerator is genuinely CP-violating. A second
charge-averaged asymmetry, defined with respect to the variables $S\equiv {\rm
sign}(\cos\theta_1\cos\theta_2)\sin\phi$ for $B_{(s)}$ and $\bar S\equiv {\rm
sign}(\cos\bar\theta_1\cos\bar\theta_2)\sin\bar\phi$ for $\bar B_{(s)}$,  is 
proportional to ${\rm Im}(A_\perp A^*_0 - \bar A_\perp\bar A^*_0)$:
\bea\label{A1cave}
{\cal A}_T^{(1) \rm chg-avg} & \equiv & \frac{[\Gamma(S>0) + \bar\Gamma(\bar
S>0)] - [\Gamma(S<0) + \bar\Gamma(\bar S<0)]} {[\Gamma(S>0) + \bar\Gamma(\bar
S>0)] + [\Gamma(S<0) + \bar\Gamma(\bar S<0)]} \nonumber\\
& = & -\frac{2\s}{\pi}\frac{{\rm Im}(A_\perp A^*_0 - \bar A_\perp\bar A^*_0)}
{(|A_0|^2 + |A_\perp|^2 + |A_\parallel|^2) + (|\bar A_0|^2 + |\bar A_\perp|^2 + 
|\bar A_\parallel|^2)}~.
\eea 

Similarly, one may define charge-averaged asymmetries
for decays in which one vector meson decays to a pseudoscalar pair 
while the other meson decays into a lepton pair. For these decays one finds
\bea\label{Alcave}
{\cal A}_T^{(2)\ell,\rm chg-avg} & = & \frac{2}{\pi}\frac{{\rm Im}
(A^\ell_\perp A^{\ell*}_\parallel - \bar A^\ell_\perp\bar 
A^{\ell*}_\parallel)}{(|A^\ell_0|^2 + |A^\ell_\perp|^2 + |A^\ell_\parallel|^2) +
 (|\bar A^\ell_0|^2 + |\bar A^\ell_\perp|^2 + |\bar A^\ell_\parallel|^2)}~,
\nonumber\\
{\cal A}_T^{(1)\ell,\rm chg-avg}
& = & \frac{\s}{\pi}\frac{{\rm Im}(A^\ell_\perp A^{\ell*}_0 -
\bar A^\ell_\perp\bar A^{\ell*}_0)} {(|A^\ell_0|^2 + |A^\ell_\perp|^2 +
|A^\ell_\parallel|^2) + (|\bar A^\ell_0|^2 + 
|\bar A^\ell_\perp|^2 + |\bar A^\ell_\parallel|^2)}~.
\eea

The two asymmetries ${\cal A}_T^{(i)\rm chg-avg}$ ($i=1,2$)
should be distinguished from somewhat different quantities 
discussed in Refs.\,\cite{Valencia:1988it,Datta:2003mj}, the average of the 
asymmetries $A^{(i)}_T$ and their charge-conjugates $\bar A^{(i)}_T$.
For instance,
\bea\label{2average}
\frac{1}{2}(A^{(2)}_T + \bar A^{(2)}_T) &\equiv&
\frac{1}{2}
 \left[\frac{\Gamma(\sin 2\phi>0)-\Gamma(\sin 2\phi<0)}
{\Gamma(\sin 2\phi>0)+\Gamma(\sin 2\phi<0)}
+ \frac{\bar\Gamma(\sin 2\bar\phi>0)-\bar\Gamma(\sin 2\bar\phi<0)}
{\bar\Gamma(\sin 2\bar\phi>0)+\bar\Gamma(\sin 2\bar\phi<0)}\right]
\nonumber\\
&=&-\frac{2}{\pi}\left (\frac{{\rm Im}(A_\perp A^*_\parallel)}{|A_0|^2 
+ |A_\perp|^2 + |A_\parallel|^2} 
- \frac{{\rm Im}(\bar A_\perp \bar A^*_\parallel)}{|\bar A_0|^2 
+ |\bar A_\perp|^2 + |\bar A_\parallel|^2} \right )~.
\eea
{\it In general this quantity is not proportional to ${\rm Im}(A_\perp
A^*_\parallel - \bar A_\perp\bar A^*_\parallel)$}.  That is, the two
asymmetries defined in Eqs.~(\ref{A2cave}) and (\ref{2average}) are 
different in the most general case. They become equal when no direct CP 
asymmetry occurs in the total decay rate, 
\beq
\Gamma(\sin 2\phi \ge 0) + \Gamma(\sin 2\phi <0) = 
\bar\Gamma(\sin 2\bar\phi \ge 0) + \bar\Gamma(\sin 2\bar\phi< 0)~,
\eeq
namely when
\beq\label{EqualRates}
|A_0|^2 + |A_\perp|^2 + |A_\parallel|^2 = 
|\bar A_0|^2 + |\bar A_\perp|^2 + |\bar A_\parallel|^2~.
\eeq

CP may be violated in decay rates for individual transversity amplitudes,
$|A_k|^2\ne
|\bar A_k|^2$ ($k=0, \parallel, \perp$). This implies nonzero CP asymmetries 
in these channels and a potential violation of (\ref{EqualRates}) leading to
$(A^{(1,2)}_T + \bar A^{(1,2)}_T)/2 \ne {\cal A}^{(1,2)\rm chg-avg}_T$.  This
happens when a given transversity amplitude obtains contributions involving
two different weak phases and two different strong phases. This
is to be contrasted with a situation involving two different weak phases, 
one associated with $A_\perp$ and the other with $A_0$ or $A_\parallel$. 
As shown in (\ref{true}) this implies a nonzero triple product asymmetry also 
when the corresponding strong phase difference vanishes. In the next subsection
we will focus on this situation in time-dependent flavorless $B_{(s)}$ decays,
assuming no direct CP violation, $|A_k|^2 = |\bar A_k|^2$ ($k=0, \parallel,
\perp$). 
 
 \subsection{Neutral $B_{(s)}$ decays to flavorless states}
 \label{subsec:flavorless}

We now consider neutral $B_{(s)}$ decays into flavorless states which are
accessible to both $B_{(s)}$ and $\bar B_{(s)}$ decays. Two examples, belonging
to the two classes considered in subsections \ref{subsec:4P} and
\ref{subsec:2P2l}, are $B_s \to \phi\phi$ and $B_s \to J/\psi\phi$.  As a
result of $B_{(s)}$-$\bar B_{(s)}$ oscillation angular decay distributions
become time-dependent.  Decay distributions for initial $B_{(s)}$ mesons are
given for these two classes by Eqs.\,(\ref{angular}) and (\ref{angular'}),
where the coefficients $|A_k|^2~(k=0,\parallel, \perp), ~{\rm Re}(A_0A^*
_\parallel),~{\rm Im}(A_\perp A^*_i)~(i=0, \parallel)$ are now functions of
time. The instantaneous transversity amplitude for a $B_{(s)}$ meson is $A_k
\equiv A_k(t=0)$. Similar expressions, in which $A_k(t)$ are replaced by $\bar
A_k(t)$, apply to angular distributions for initial $\bar B_{(s)}$ mesons with
$\bar A_k \equiv \bar A_k(t=0)$. In particular, terms relevant for triple
products involving ${\rm Im}[A_\perp(t) A^*_i(t)]$ and ${\rm Im}[\bar A_\perp(t)
\bar A^*_i(t)]$ appear with {\it equal signs} in the distributions for initial
$B_{(s)}$ and $\bar B_{(s)}$.  Thus, time-dependent TP quantities measured in
untagged neutral $B_{(s)}$ decays to flavorless states are of the form ${\rm Im}
[A_\perp(t) A^*_i(t) + \bar A_\perp(t) \bar A^*_i(t)]$.  [Note that the
corresponding time-independent terms in Eqs.\,(\ref{angular}) and
(\ref{angularBbar}) appear with opposite signs for two distributions written in
terms of $\theta_1, \theta_2, \phi$ and $\bar\theta_1, \bar\theta_2, \bar\phi$.
The opposite relative signs in the two cases may be explained by noting that
interchanging final particle and antiparticle momenta in decays to flavorless
states corresponds to $\sin\bar\phi = - \sin\phi$ while the functions of
$\theta_i$ and $\bar\theta_i$ are equal.]

Let us study flavor-untagged decays which involve the time-dependent triple
products ${\rm Im}[A_\perp(t) A^*_i(t) + \bar A_\perp(t) \bar A^*_i(t)]~(i=0,
\parallel)$. Considering their values at $t=0$, ${\rm Im}(A_\perp A^*_i + \bar
A_\perp\bar A^*_i)$, we now show that these two quantities are genuinely CP
violating.  Using standard notations for $B_{(s)}$-$\bar B_{(s)}$ mixing and 
assuming no CP violation in mixing ($|q/p|=1$) and decay ($|\bar A_k|=|A_k|$), 
one has~\cite{Branco:1999fs}
\beq\label{phik}
\frac{q}{p}\frac{\bar A_k}{A_k} = \eta_k e^{-2i\phi_k}~.
\eeq
Here $\eta_k$ is the CP parity for a state of transversity $k$ ($\eta_0 =
\eta_\parallel= -\eta_\perp =+1$), while $\phi_k$ is the weak phase involved in
an interference between mixing and decay amplitudes. Denoting the CP conserving
strong phase of $A_k$ by $\delta_k$, $A_k=|A_k|e^{i\delta_k}e^{i\phi_k}$,
so $\bar A_k = (p/q) \eta_k e^{i \delta_k} e^{-i \phi_k}$,
one has for $i=0, \parallel$:
\bea\label{Im}
{\rm Im}(A_\perp A^*_i + \bar A_\perp\bar A^*_i) & = & |A_\perp| |A_i|{\rm Im}
\left [e^{i(\delta_\perp - \delta_i)}(e^{i(\phi_\perp-\phi_i)}
- e^{-i(\phi_\perp-\phi_i)})\right] \nonumber\\
& = & 2|A_\perp| |A_i|\cos(\delta_\perp - \delta_i) \sin(\phi_\perp-\phi_i)~.
\eea
As argued above, this ``true" CP violating quantity is nonzero also when the 
CP-conserving phase difference vanishes, provided that the CP-violating 
phase difference between the two transversity amplitudes is nonzero.  Note the
change of relative sign between terms on the left-hand-side of Eqs.
(\ref{true}) and (\ref{Im}), defining ``true"
CP violating asymmetries in decays into specific flavor states and into 
flavorless CP states of opposite CP parity, respectively.

Time-dependence of the CP violating triple products ${\rm Im}[A_\perp(t)
A^*_i(t) + \bar A_\perp(t) \bar A^*_i(t)]$ $(i=0, \parallel)$ depends on the
$B_{(s)}$-$\bar B_{(s)}$ oscillation frequency determined by a mass difference
$\Delta m$ and on a width difference $\Delta\Gamma$ affecting the exponential 
decay. Early studies of time-dependent angular distributions
\cite{Dunietz:1995cp}, applied in particular to $B_s\to J/\psi\phi$, have
assumed that a single weak phase, common to all three transversity states, is 
associated with interference between $B_s$-$\bar B_s$ mixing and decay
amplitudes.  In this case ($\phi_\perp=\phi_i$) the above two triple products
vanish. 
Refs.~\cite{Valencia:1988it,Datta:2003mj} study some aspects of TP asymmetries
induced by $B$-$\bar B$ mixing.
We will now generalize the time-dependence of the two triple products
to the case under consideration, $\phi_\perp \ne \phi_i~(i=0,\parallel)$. Our
calculation applies to both strange and nonstrange neutral mesons, $B=B^0, B_s$
and their antiparticles, $\bar B=\bar B^0, \bar B_s$. 

One starts with evolution equations for $B$ and $\bar B$~\cite{Branco:1999fs}
\beq
B_(t) = g_+(t)B + (q/p)g_-(t)\bar B~,~~~~~~~
\bar B(t) = (p/q)g_-(t)B + g_+(t)\bar B~.
\eeq
where
\bea
g_+(t) & = & e^{-imt}e^{-\Gamma t/2}[\cosh(\Delta\Gamma t/4)\cos(\Delta mt/2) - 
i\sinh(\Delta\Gamma t/4)\sin(\Delta mt/2)],\nonumber\\
g_-(t) & = & e^{-imt}e^{-\gamma t/2}[-\sinh(\Delta\Gamma t/4)\cos(\Delta mt/2) +
i\cosh(\Delta\Gamma t/4)\sin(\Delta mt/2)],
\eea
\bea
|g_\pm(t)|^2 &=& (e^{-\Gamma t}/2)[\cosh(\Delta\Gamma t/2) \pm \cos(\Delta mt)]~,
\nonumber\\
g^*_+(t)g_-(t) &=& (e^{-\Gamma t}/2)[-\sinh(\Delta\Gamma t/2) + i\sin(\Delta mt)]~.
\label{g^2}
\eea
Time dependence of transversity amplitudes, $A_k\equiv \langle k|B\rangle,
\bar A_k\equiv \langle k|\bar B\rangle$~($k=0, \parallel, \perp$), is given by:
\bea
A_k(t) &\equiv & \langle k | B(t) \rangle = g_+(t) A_k +(q/p) g_-(t) \bar A_k~,
\nonumber\\
\bar A_k(t) & \equiv & \langle k | \bar B(t) \rangle = (p/q)g_-(t)A_k + g_+(t)\bar A_k~.
\eea

We are interested in interference terms $A^*_i(t)A_k(t)$ and $ \bar A^*_i(t)\bar A_k(t)$.
Using Eqs. (\ref{phik}) and (\ref{g^2}) one obtains
\bea
&& A^*_i(t)A_k(t) = [g^*_+A^*_i + (q/p)^*g^*_-\bar A^*_i][g_+A_k + (q/p)g_-\bar A_k]
\nonumber\\
&&=A^*_iA_k[|g_+|^2 + (q/p)(\bar A_k/A_k)g^*_+g_-]
+ \bar A^*_i\bar A_k[|g_-|^2 + (p/q)(A_k/\bar A_k)g_+g^*_-]
\nonumber \\
&& = \frac{e^{-\Gamma t}}{2}\left [A^*_iA_k \left( \cosh(\Delta\Gamma t/2) + \cos(\Delta mt)
+\eta_ke^{-2i\phi_k} [-\sinh(\Delta\Gamma t/2) + i\sin(\Delta mt)] \right) \right.
\nonumber\\
&&~~~~~~~~\left.+\bar A^*_i\bar A_k \left( \cosh(\Delta\Gamma t/2) - \cos(\Delta mt)
+\eta_ke^{2i\phi_k}[-\sinh(\Delta\Gamma t/2) - i\sin(\Delta mt)] \right) \right ].
\nonumber\\
\eea
Inserting $A^*_iA_k = |A_i||A_k|e^{i(\delta_k-\delta_i)}e^{i(\phi_k-\phi_i)}$,
$\bar A^*_i\bar A_k =  \eta_i\eta_k|A_i||A_k|e^{i(\delta_k-\delta_i)}e^{-i(\phi_k-\phi_i)}$,
implies for $i=0, \parallel, k=\perp$,
\bea
&& A^*_i(t)A_\perp(t) = e^{-\Gamma t} |A_i||A_\perp|e^{i(\delta_\perp-\delta_i)}
\left[i\sin(\phi_\perp-\phi_i)\cosh(\Delta\Gamma t/2) +\cos(\phi_\perp-\phi_i)
\cos(\Delta mt) \right.
\nonumber\\
&&~~~~~~~~~~~~~~~~~~
~~~~~~~~~~~~~~~ \left.-~i\sin(\phi_\perp+\phi_i)\sinh(\Delta\Gamma t/2)  
-i\cos(\phi_\perp+\phi_i)\sin(\Delta mt) \right],
\eea
leading to
\bea
&&{\rm Im}[A^*_i(t)A_\perp(t)]
\nonumber\\
&&= e^{-\Gamma t} |A_i||A_\perp| \left( \cos(\delta_\perp-\delta_i)
[ \sin(\phi_\perp-\phi_i)\cosh(\Delta\Gamma t/2)
-\sin(\phi_\perp+\phi_i)\sinh(\Delta\Gamma t/2)   
\right.
\nonumber \\
&&~~~~~~~~~~~~~~~ \left. - \cos(\phi_\perp+\phi_i)\sin(\Delta mt)] 
+ \sin(\delta_\perp-\delta_i)\cos(\phi_\perp-\phi_i)\cos(\Delta mt) \right)~.
\eea
Similarly one has 
\bea
&& {\rm Im}[\bar A^*_i(t)\bar A_\perp(t)]
\nonumber\\
&&= e^{-\Gamma t} |A_i||A_\perp| \left( \cos(\delta_\perp-\delta_i)
\left[ \sin(\phi_\perp-\phi_i)\cosh(\Delta\Gamma t/2)
-\sin(\phi_\perp+\phi_i)\sinh(\Delta\Gamma t/2)   
\right.\right.
\nonumber \\
&&~~~~~~~~~~~~~~~ \left.\left. + \cos(\phi_\perp+\phi_i)\sin(\Delta mt) \right ] 
- \sin(\delta_\perp-\delta_i)\cos(\phi_\perp-\phi_i)\cos(\Delta mt) \right)~.
\eea
Thus
\bea\label{untagged}
&&{\rm Im}[A_\perp(t) A^*_i(t) + \bar A_\perp(t) \bar A^*_i(t)] 
=  2|A_\perp||A_i|e^{-\Gamma t}\cos(\delta_\perp - \delta_i)
\nonumber\\
&&~~~~~~~ \left [\sin(\phi_\perp -\phi_i)\cosh(\Delta\Gamma t/2) 
- \sin(\phi_\perp+\phi_i)\sinh(\Delta\Gamma t/2)\right]~.
\eea
This time-dependent result agrees with (\ref{Im}) at $t=0$.  It demonstrates
for arbitrary time a behavior of a genuine CP violating quantity
which does not vanish for nonzero weak phases and requires no strong phases.

The two ``true" CP violating time-integrated triple product asymmetries 
($i=0, \parallel$) for untagged decays are proportional to  
\bea\label{int}
&& \Gamma \int_0^\infty {\rm Im}[A_\perp(t) A^*_i(t) + \bar A_\perp(t)
\bar A^*_i(t)]dt =  2|A_\perp||A_i|\cos(\delta_\perp - \delta_i)
\nonumber\\
&&~~~~~~\left (\sin(\phi_\perp -\phi_i)
- \sin(\phi_\perp+\phi_i)(\Delta\Gamma/2\Gamma) + 
{\cal O}[(\Delta\Gamma/2\Gamma)^2]\right )~.
\label{trueint}
\eea
We conclude that sizable CP violating TP asymmetries do not require direct CP
violation. They do require however that the weak phases dominating
$A_i ~(i=0,\parallel)$ and $A_\perp$ differ from one another.  In the special
case $\phi_\perp =\phi_i$ considered in Ref.\ \cite{Dunietz:1995cp} (including
the Standard Model) the first term in (\ref{int}) vanishes while the remaining
term is suppressed by $\Delta\Gamma/2\Gamma$.  

It is interesting (and perhaps surprising) that the time-integrated asymmetries
for untagged $B_s$ decays are not suppressed due to fast $B_s$-$\bar B_s$
oscillations by $(\Gamma_s/\Delta
m_s)^2$ or by $\Gamma_s/\Delta m_s$, as they would be for time-dependent terms
behaving like $\cos(\Delta mt)$ or $\sin(\Delta mt)$. This behavior
characterizes the two ``fake" asymmetries which are proportional to
\bea
&&{\rm Im}[A_\perp(t) A^*_i(t) - \bar A_\perp(t) \bar A^*_i(t)] 
=  2|A_\perp||A_i|e^{-\Gamma t}
\nonumber\\
&&~\left[\sin(\delta_\perp-\delta_i)\cos(\phi_\perp-\phi_i)\cos(\Delta mt)
-\cos(\delta_\perp-\delta_i) \cos(\phi_\perp+\phi_i)\sin(\Delta mt) \right].
\eea
For $B_s$ decays the corresponding time-integrated fake asymmetries are 
suppressed by powers of $\Gamma_s/\Delta m_s\sim 0.04$~\cite{Barberio:2008fa}:
\bea
&& \Gamma \int_0^\infty {\rm Im}[A_\perp(t) A^*_i(t) - \bar A_\perp(t)
\bar A^*_i(t)]dt \approx  2|A_\perp||A_i|
\nonumber\\
&& ~\left[\sin(\delta_\perp-\delta_i)\cos(\phi_\perp-\phi_i)(\Gamma_s/\Delta m_s)^2
-\cos(\delta_\perp-\delta_i) \cos(\phi_\perp+\phi_i)\Gamma_s/\Delta m_s \right].
\label{fakeint}
\eea
Note that measurements of both time-dependent and time-integrated fake
asymmetries do require flavor tagging. 

\section{Triple products in specific $B_{(s)} \to V_1 V_2$ decays
\label{sec:TP2V}}

The first class of decays we shall discuss in this section includes processes
dominated by a penguin
$b \to s$ amplitude.  Before treating asymmetries associated with specific
final states it is worth noting polarization properties in such decays.
We shall then discuss TP asymmetries in $B\to \phi K^*$ and $B_s \to \phi
\phi$.

\subsection{Polarization in penguin-dominated decays
\label{subsec:pol}}

We shall reiterate a discussion given in Ref.\ \cite{Rosner:2011ux}.
The decays $B \to \phi K^*$ and $B_s \to \phi \phi$ are both dominated by the
$b \to s$ penguin diagram.  Factorization predicts dominant longitudinal
polarization of the vector mesons, in contrast to observations
\cite{CDF10120,Ba99,Ba97}.  Table \ref{tab:fact}  quotes longitudinal and
transverse fractions for the above
penguin-dominated processes as well as for $B^{(+,0)}\to\rho^0 K^{*(0,+)}$
which belong to the same class.  By contrast, the
tree-dominated decay $B^0 \to \rho^+ \rho^-$ has $f_L= 0.992 \pm 0.024^{+0.026}
_{-0.013}$ \cite{Aubert:2007nua}, or nearly 1 as predicted.  There is no reason
to trust factorization for the penguin amplitude, which may be due to
rescattering from charm-anticharm intermediate states.  Although $f_L < 1$ in
penguin-dominated decays has frequently been quoted as possible evidence for
new physics (see, e.g., \cite{Datta:2011qz};
however see also~\cite{Beneke:2006hg}), we prefer to reserve judgment
on this issue.

\begin{table}
\caption{Longitudinal and transverse fractions $f_L$ and $f_T$ for some
$b\to s$-penguin $B \to VV$ processes.
\label{tab:fact}}
\begin{center}
\begin{tabular}{c c c c c} \hline \hline
      & $B_s \to \phi \phi$ & $B^+ \to \phi K^{*+}$ & $B^+ \to \rho^0 K^{*+}$
 & $B^0 \to \rho^0 K^{*0}$ \\
 & \cite{CDF10120} & \cite{Ba99} & \cite{Ba97} & \cite{Ba97} \\ \hline
$f_L$ & 0.348$\pm$0.041$\pm$0.021 & 0.49$\pm$0.05$\pm$0.03 &
 0.52$\pm$0.10$\pm$0.04 & 0.57$\pm$0.09$\pm$0.08 \\
$f_T$ & 0.652$\pm$0.041$\pm$0.021 & 0.51$\pm$0.05$\pm$0.03 &
 0.48$\pm$0.10$\pm$0.04 & 0.43$\pm$0.09$\pm$0.08 \\ \hline \hline
\end{tabular}
\end{center}
\end{table}

\subsection{$B\to \phi K^*$
\label{subsec:phiK}}

True and fake TP asymmetries were defined in subsection~\ref{subsec:self-tag} as
\beq
{\cal A}_T^{\rm true} \propto {\rm Im}(A_\perp A_i^* - \bar A_\perp \bar
A_i^*)~,~~ {\cal A}_T^{\rm fake} \propto {\rm Im}(A_\perp A_i^* + \bar
A_\perp \bar A_i^*)~,~~(i=0,\parallel)~,
\eeq
using normalizations for ${\cal A}^{(1)}_T$ and ${\cal A}^{(2)}_T$ as in 
the second line of Eqs. (\ref{A1cave}) and (\ref{A2cave}). 
From $B^0 \to \phi K^{*0}$ amplitudes 
and relative phases quoted in~\cite{Barberio:2008fa}
 we estimate
\beq
A_T^{(1)}=-0.117\pm 0.022;
\bar A_T^{(1)}= +0.091\pm 0.023;
 A_T^{(2)}=-0.003\pm 0.045;
 \bar A_T^{(2)}=-0.006\pm 0.041~.
\eeq
These values imply a large fake ${\cal A_T}^{(1)}$ (since $A_T^{(1)} - \bar
A_T^{(1)} \ne 0$); no true ${\cal A}_T^{(1)}$ (since $A_T^{(1)} + \bar
A_T^{(1)}$ is consistent with zero); and no fake {\it or} true
${\cal A}_T^{(2)}$ (since both $ A_T^{(2)}$ and $\bar A_T^{(2)}$ are consistent
with zero).  The large fake ${\cal A}_T^{(1)}$ simply
reflects the importance of strong final-state phases.

\subsection{$B_s \to \phi\phi$}
\label{subsec:phiphi}

True triple product asymmetries discussed in subsection~\ref{subsec:flavorless}
with definitions as in the first line of
Eqs.~(\ref{A2cave}) and (\ref{A1cave}) are
related to those recently reported by Dorigo on behalf of the CDF Collaboration
for the decay $B_s \to \phi \phi$ \cite{Dorigo:2011ef}. The measured values are 
${\cal A}_u \leftrightarrow {\cal A}_T^{(2)}= (-0.7 \pm 6.4 \pm 1.8)\%$; 
${\cal A}_v \leftrightarrow {\cal A_T}^{(1)} = (-12.0 \pm 6.4 \pm 1.6)\%$.
These observables represent time-integrated and untagged quantities, to which
Eq.\ (\ref{trueint}) applies.
As mentioned, these two triple product asymmetries require
non-zero values of the weak phase differences $\phi_\perp - \phi_\parallel$ 
and $\phi_\perp - \phi_0$, respectively, to avoid
being suppressed by a factor of $\Delta \Gamma_s / 2 \Gamma_s < 0.1$
\cite{Lenz:2011ti}.

\subsection{$B_s\to J/\psi\phi$}\label{Jphi}

Angular and time-dependence studied for $B_s\to J/\psi\phi$ by the
CDF~\cite{Aaltonen:2007he} and D0~\cite{:2008fj} collaborations  provided 
information on
the weak phase occurring in the interference between $B_s$-$\bar B_s$ mixing 
and $b\to c\bar c s$ decay. This phase, expected to be very small in the CKM 
framework~\cite{Barberio:2008fa}, may obtain corrections from new physics 
contributions to $B_s$-$\bar B_s$ mixing. Here we are interested in lessons to 
be learned from measuring CP violating triple product asymmetries in this
process.
  
Triple product asymmetries in this class of decays were studied in Section 
\ref{subsec:2P2l} in terms of tranversity amplitudes. 
Time-dependent CP violating asymmetries given by Eq.~(\ref{untagged}) are 
obtained by adding up events for initial $B_s$ and initial $\bar B_s$.
The first term, $\propto \sin(\phi_\perp -\phi_i)\cosh(\Delta\Gamma_s t/2)$
($i=0, \parallel$), vanishes for $\phi_\perp=\phi_i$, while the second term,
$\propto -\sin(\phi_\perp + \phi_i)\sinh(\Delta\Gamma_s t/2)$, remains nonzero
in this limit. The phases $\phi_k$ ($k=0, \parallel, \perp$), occurring in the
interference of the mixing amplitude with the three transversity amplitudes
[see Eq.~(\ref{phik})], are equal in the CKM framework. They are expected to be
equal to a very good approximation also in extensions of this framework because
$b\to c\bar cs$ is CKM-favored. The quantity which can potentially be affected
in new physics schemes is $\phi_\perp + \phi_i \approx 2\phi_k$ ($k=0,
\parallel, \perp$) which determines the magnitude 
of the coefficient of the $\sinh(\Delta\Gamma_s t/2)$ term in the CP-violating TP 
asymmetry. This coefficient is of order a few percent in the CKM framework but 
may be sizable in the presence of new contributions to $B_s$-$\bar B_s$
mixing. This term is suppressed by
$\Delta\Gamma_s/2\Gamma_s$ when time-integrated.

\section{Concluding remarks \label{sec:concl}}

We have discussed the differences between ``true'' CP-violating triple
product (TP) asymmetries which require no strong phases, and ``fake''
asymmetries which require non-zero strong phases but no CP violation.
We have shown that TP asymmetries vanish for two identical and kinematically
indistinguishable particles in the final state, demonstrating this property
through two examples of Cabibbo-favored four-body $D$ decays. 
Such asymmetries need not vanish even when two identical particles are
present as long as they have non-trivial kinematic correlations, as in
$K_L \to e^+ e^- e^+ e^-$.
We have shown that while triple product asymmetries in charmed meson 
decays do not manifest CP violation, they display an interesting pattern
of final-state interactions correlated with total decay widths.

We studied TP asymmetries in $B$ and $B_s$ meson decays to two vector mesons
each decaying to a pseudoscalar pair, extending results to decays where one
vector meson decays into a lepton pair. We derived expressions for
time-dependent TP asymmetries for neutral $B$ and $B_s$ decays to flavorless
states in terms of the neutral $B_{(s)}$
mass difference $\Delta m$ and the width-difference $\Delta\Gamma$.
Time-integrated true CP violating asymmetries, measurable for untagged $B_s$
decays, were shown to be suppressed by neither $\Gamma_s/\Delta m_s$ nor
$\Delta\Gamma_s/\Gamma_s$, but to require two different weak phases in
decays to CP-even and CP-odd transversity states.
Finally, implications were discussed for TP  asymmetries in $B\to K^*\phi, 
B_s \to \phi\phi$ and $B_s \to J/\psi\phi$.

\bigskip

M. G. is grateful to the Enrico Fermi Institute at the University of Chicago
and to the CERN Theory Division
for their kind hospitality.  We thank Mirco Dorigo and Diego Tonelli for
helpful questions, and Boris Kayser, Alakabha Datta, David London,
Brian Meadows and Maurizio Martinelli for useful
communications.  This work was supported in part by the United States
Department of Energy through Grant No.\ DE FG02 90ER40560.

\end{document}